\documentclass[11pt,a4paper]{article}
\usepackage[dvips]{graphics}
\makeatletter \oddsidemargin -.1in \evensidemargin -.1in
\newcommand{\singlespacing}{\let\CS=\@currsize\renewcommand{\baselinestretch}{1}\tiny\CS}
\newcommand{\doublespacing}{\let\CS=\@currsize\renewcommand{\baselinestretch}{1.35}\tiny\CS}
\usepackage{graphics}
\usepackage{amsmath}
\usepackage{amssymb}
\usepackage{float, epsfig, floatflt, here}
\usepackage{cite}
\usepackage{longtable}
\usepackage{graphicx}
\usepackage{pdfpages}
\usepackage{makeidx}
\usepackage{caption}
\usepackage{subcaption}
\usepackage[export]{adjustbox}
\usepackage{quantikz}
\usepackage{xcolor}
\title {\textbf{\LARGE   Hybrid Multi-Directional Quantum Communication Protocol}}
\author{ { Mitali Sisodia}$^{1}$ \thanks{e-mail: mitalisisodiyadc@gmail.com}, {Manoj Kumar Mandal}$^{2}$ \thanks{e-mail: manojmandaliiest@gmail.com}, { Binayak S. Choudhury}$^{2}$ \thanks {e-mail : binayak@math.iiests.ac.in}\\
~\\
$^{1}$  Department of Physics,
 Indian Institute of Technology,  Delhi\\
 New Delhi 110016, India\\
$^{2}$ Department of Mathematics,
 Indian Institute of Engineering Science and Technology,\\ Shibpur
 B. Garden, Howrah - 711103,
 West Bengal, India\\}
 
\begin{document}
\date{}\maketitle

\begin{abstract}
The way a new type of state called a hybrid state, which contains more than one degree of freedom, is used in many practical applications of quantum communication tasks with lesser amount of resources. Similarly, our aim is here to perform multi quantum communication tasks in a protocol to approach quantum information in multipurpose and multi-directional. We propose a hybrid multi-directional six-party scheme of implementing quantum teleportation and joint remote state preparation under the supervision of a controller via a multi-qubit entangled state as a quantum channel with $100\%$ success probability. Moreover, we analytically derive the average fidelities of this hybrid scheme under the amplitude-damping and the phase-damping noise. 
\end{abstract}

\section{Introduction}
One of the most surprising consequences in quantum information science is the quantum teleportation (QT) technique, which has been increasingly investigated as a basic element for various applications in quantum technology. The first teleportation scheme was proposed by Bennett et al. in 1993, in which a two-qubit entangled state (Bell state) has been utilized as a quantum resource to teleport a single-qubit unknown quantum state \cite{teleportation}. The essence of teleportation is that the state to be teleported is unknown to the sender. Later, a separate group of protocols known as remote state preparation (RSP) \cite{RSP} was developed in which the state to be transferred is known, joint remote state preparation (JRSP) \cite{JRSP} was advanced in which information of the state to be transferred is divided amongst more than one party. Protocols for the purpose of performing modified and highly efficient quantum communication tasks have been subsequently discussed in several works, such as bidirectional teleportation \cite{b3,b4,b5}, quantum conferencing \cite{c1,c2}, controlled quantum teleportation (CQT) \cite{cm1,cm2,cm3}. \\ 
QT is a one-way quantum communication process in which an unknown quantum state is teleported by the sender to the receiver with the use of an entangled quantum resource. A few years later, variants of QT have been studied which predict that simultaneous two-way communication is possible with the use of double QT resource, known as bidirectional quantum teleportation (BQT). Further, a three-party QT scheme, i.e., CQT, has been proposed with the use of a three-qubit entangled state. In which, a party controls or supervises the roles of different involved parties and helps to perform the teleportation task between sender and receiver. So far, a single quantum communication task has been performed in a protocol by using a number of entangled quantum resources with a single degree of freedom. Later, it was studied that multi-quantum communication tasks can be performed in a protocol, which is known as hybrid communication and the number of degrees of freedom can be used in a quantum state, i.e., hybrid state, to increase the practical applications in quantum information processing \cite{hs1,hs2,hs3,hs4}.\\
Among multitasking protocols, hybrid protocols are the multitasking communication processes where more than one type of quantum protocols are combined to form an integrated process discussed in \cite{h3,h4,h5,h6,h7,h8}. For instance, a hybrid protocol in which four-party controlled JRSP and CQT has been implemented via a seven-qubit entangled state \cite{h3}. Group of Gong et al. reported the bidirectional hybrid protocol \cite{h4} and multi-party controlled cyclic hybrid quantum communication protocol \cite{h9} under the effect of different types of noises (bit flip, phase flip, bit-phase flip, phase-damping and amplitude-damping noise): the work of Joo et al. developed the hybrid scheme that teleports quantum information from a solid-state qubit to microwave photonic state \cite{h8}.  In Ref. \cite{h9}, a hybrid (QT and RSP scheme) double-channel scheme is proposed and also checked the effect of noises on the scheme. Recently, a hierarchical controlled hybrid quantum communication scheme has been presented for specific IoT (Internet of Things) application scenarios and simulated the results on IBM’s Qiskit Aer quantum computing simulator \cite{h10} and another hybrid scheme which is based on a classical-quantum communications protocols for managing classical blockchains is studied by Liu A. et al. in 2023 \cite{h11}. Since then,  rapid development in hybrid quantum communication has attracted much attention and motivation from researchers to move forward into it.

By motivating and recognizing the several applications and advantages of hybrid communication, in the present context, a multitasking hybrid protocol under the supervision of a controller is presented where three single-qubit states are transferred through a JRSP-type process while the receivers of the three single-qubit states teleport to the sender one-qubit, two-qubit, and three-qubit states respectively. Also, we assume the distribution of qubits among the parties in an open quantum system (interaction with the noisy environment) \cite{n1, adpd, n2, n3}, where quantum noise will unavoidably affect the qubits. As a result, the pure state becomes mixed, which has a significant impact on the scheme's efficiency (or the integrity of the output state). Keeping this point in mind, in this context, we also study the effect of two Markovian noisy channels amplitude-damping (AD) and phase-damping (PD) on this hybrid scheme by calculating the average fidelities.

The paper is organized as follows: in Sec. \ref{sec:Hybrid-Protocol}, a complete hybrid multi-directional CQT protocol is discussed. In Sec. \ref{sec:noise}, the protocol is considered in noisy environments and AD and PD noise is discussed in the subsections. Finally, the work is concluded in the Sec. \ref{sec:conclusion}.\\
 
\section{Hybrid Multi-directional Controlled Quantum communication(HMCQC) Protocol \label{sec:Hybrid-Protocol}}
In this HMCQC protocol, we consider six parties: Alice, Bob, Charlie, Davis, Candy, and Simon. Alice and Charlie want to create jointly three single-qubit known states in the hands of Bob, Davis, and Candy separately, and at the same time, each party Bob, David, and Candy want to teleport three different qubit states to Alice under the controller Simon.
Suppose, Alice and Charlie have three known single-qubit state $|\phi_0\rangle$, $|\phi_1\rangle$, and $|\phi_2\rangle$  which are given by
 \begin{equation}
     \begin{split}
         |\phi_0\rangle=&x_0|0\rangle+y_0e^{i\theta_0}|1\rangle\\
         |\phi_1\rangle=&x_1|0\rangle+y_1e^{i\theta_1}|1\rangle\\
         |\phi_2\rangle=&x_2|0\rangle+y_2e^{i\theta_2}|1\rangle.
     \end{split}
 \end{equation}
where  $x_i$, $y_i$, and $\theta_i$ (where $i=0,1,2$) are real numbers and $\theta_j$ (where $j=0,1,2$) is  in the range $[0, 2\pi]$. Alice knows the value of $x_i$ and $y_i$, while Charlie knows the values of $\theta_j$. Additionally, the real numbers $x_i$ and $y_i$ satisfy the normalization condition $x_i^2 + y_i^2 = 1$.\\

 The state $|\phi_0\rangle$ is to be created at Bob's place, the state $|\phi_1\rangle$ is to be created at David's place and the state $|\phi_2\rangle$ is to be created at Candy's place.\\

 At the same time, Bob, David, and Candy want to teleport three unknown states $|\psi_0\rangle$, $|\psi_1\rangle$, and $|\psi_2\rangle$ respectively to Alice. The unknown quantum states are given by
 \begin{equation}\label{eq2}
     \begin{split}
         |\psi_0\rangle=&(a_0|0\rangle+b_0|1\rangle)_{1}\\
         |\psi_1\rangle=&(a_1|00\rangle+b_1|11\rangle)_{23}\\
         |\psi_2\rangle=&(a_2|000\rangle+b_2|111\rangle)_{456}.
     \end{split}
 \end{equation}
 where  $a_i$ and $b_i$ (where $i=0,1,2$) are complex numbers satisfying the condition $|a_i|^2 + |b_i|^2 = 1$. These coefficients are unknown to all parties.\\
 
\begin{figure}
        \centering
        \includegraphics[width=0.8\linewidth]{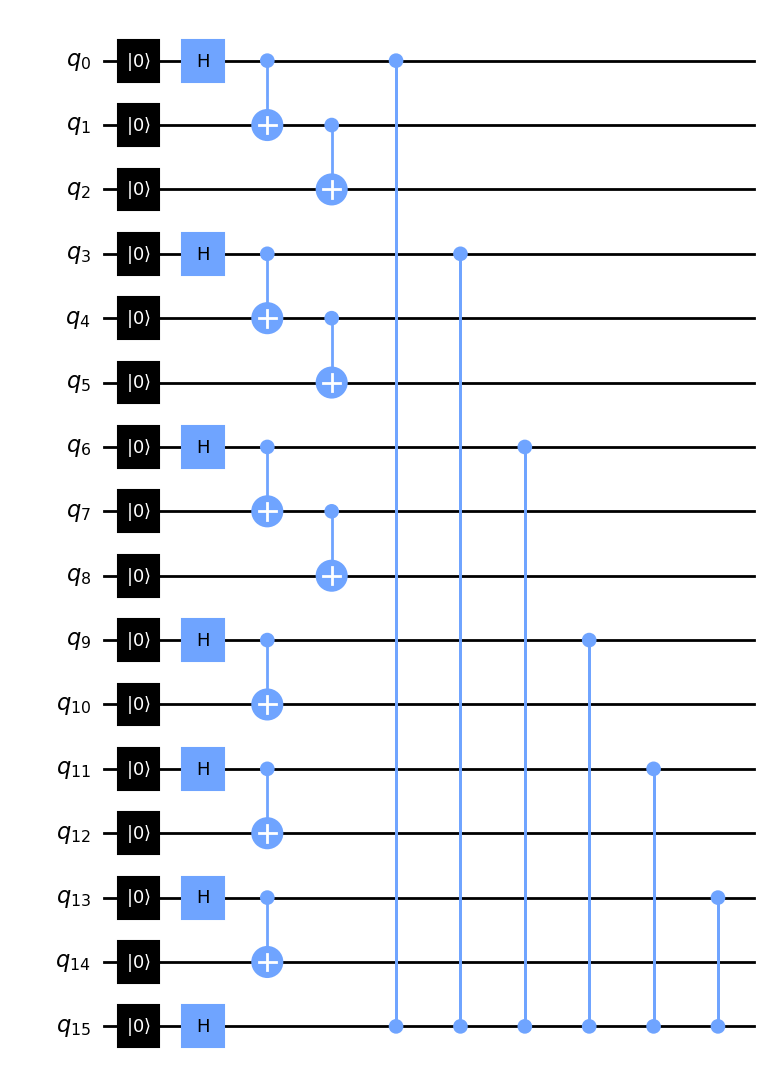}
        \caption{Quantum circuit for the generation of the entangled state $|\tau\rangle$ in equation (\ref{eq3}). }
        \label{fig:1}
    \end{figure}

At first, David applies the CNOT gate on his unknown two-qubit state $(2,3)$ where qubit $2$ is the controlled qubit and qubit $3$  is the target qubit. Candy also applies two CNOT gates. The first CNOT gate is on qubit pair $(4,5)$ with qubit $4$ as the controlled qubit and qubit $5$ as the target qubit. Second, another CNOT gate is applied on qubit pair $(5,6)$ with qubit $5$ as the controlled qubit and qubit $6$ as the target qubit. Then the state of the qubits given in equations $(\ref{eq2})$ is transformed into 
\begin{equation}
     \begin{split}
         |\psi_0'\rangle=&(a_0|0\rangle+b_0|1\rangle)_{1}\\
         |\psi_1'\rangle=&(a_1|0\rangle+b_1|1\rangle)_{2}\otimes|0\rangle_{3}\\
         |\psi_2'\rangle=&(a_2|0\rangle+b_2|1\rangle)_{4}\otimes|00\rangle_{56}.
     \end{split}
 \end{equation}
 Now Bob, David and Candy will teleport only qubits $1$, $2$ and $4$ to Alice, respectively.
 To achieve this protocol, all the parties shared a 16-qubit entangled quantum state as a quantum channel, which is given as
 \begin{equation}\label{eq3}
     \begin{split}
         |\tau\rangle=&\frac{1}{\sqrt{2}}|W^+\rangle_{A_0C_0B_0}\otimes|W^+\rangle_{A_1C_1D_0}\otimes|W^+\rangle_{A_2C_2E_0}\otimes|\Phi_{00}\rangle_{B_1A_3}\otimes|\Phi_{00}\rangle_{D_1A_4}\otimes|\Phi_{00}\rangle_{E_1A_5}\otimes|0\rangle_{S}\\
         +&\frac{1}{\sqrt{2}}|W^-\rangle_{A_0C_0B_0}\otimes|W^-\rangle_{A_1C_1D_0}\otimes|W^-\rangle_{A_2C_2E_0}\otimes|\Phi_{01}\rangle_{B_1A_3}\otimes|\Phi_{01}\rangle_{D_1A_4}\otimes|\Phi_{01}\rangle_{E_1A_5}\otimes|1\rangle_{S}.
     \end{split}
 \end{equation}
 where the qubits $A_0, A_1, A_2, A_3, A_4$ and $A_5$ belong to Alice, qubits $C_0, C_1$ and $C_2$ belong to Charlie,  $B_0$ and $B_1$ are in the hands of Bob,  qubits $D_0$ and $D_1$ belong to David and qubits $E_0$ and $E_1$ belong to Candy while qubit $S$ belongs to the controller Simon.  Quantum circuit for the generation of the entangled state given in Equation (4) is given in Figure \ref{fig:1}.  Also the states $|W^{\pm}\rangle=\frac{1}{\sqrt{2}}(|000\rangle\pm|111\rangle)$, $|\Phi_{00}\rangle=\frac{1}{\sqrt{2}}(|00\rangle+|11\rangle)$ and $|\Phi_{01}\rangle=\frac{1}{\sqrt{2}}(|00\rangle-|11\rangle)$.\\

\noindent The initial quantum state of the entire system is given by
\begin{equation}
    |X\rangle=(a_0|0\rangle+b_0|1\rangle)_{1}\otimes(a_1|0\rangle+b_1|1\rangle)_{2}\otimes(a_2|0\rangle+b_2|1\rangle)_{4}\otimes |\tau\rangle_{A_0C_0B_0, \cdot \cdot \cdot, S},
\end{equation}
 The entire process of the HMCQC protocol is described as follows.\\
 
 \noindent\textbf{Step 1:} Alice makes  single-qubit measurement on her qubits $A_0, A_1$ and $A_2$ with the X-basis $\{|\alpha_0^0\rangle, |\alpha_1^0\rangle\}$, $\{|\alpha_0^1\rangle, |\alpha_1^1\rangle\}$ 
 and $\{|\alpha_0^2\rangle, |\alpha_1^2\rangle\}$ respectively, which are given by
 \begin{equation}
     \begin{split}
         |\alpha_0^j\rangle=\frac{1}{\sqrt{2}}(x_j|0\rangle+y_j|1\rangle) ~~~(where~~ j=0,1,2)\\
         |\alpha_1^j\rangle=\frac{1}{\sqrt{2}}(y_j|0\rangle-x_j|1\rangle),~~~(where~~ j=0,1,2)
     \end{split}
 \end{equation}
 After the measurement, she shares her measurement results $\{|\alpha_l^j\rangle: l=0,1; j=0,1,2\}$ with Charlie classically. According to the measurement result of Alice, Charlie chooses measurement bases  for measuring his qubits $C_0, C_1$ and $C_2$.\\
 
\noindent When Alice's measurement results are $|\alpha_0^j\rangle_{A_j}: j=0,1,2$ then Charlie choose his measurement bases $\{|\beta_0^{0,j}\rangle, |\beta_1^{0,j}\rangle\}: j=0,1,2$ for measuring qubits $C_j:j=0,1,2$ respectively, which are given as
 \begin{equation}
     \begin{split}
         |\beta_0^{0,j}\rangle=\frac{1}{\sqrt{2}}(|0\rangle+e^{-i\theta_j}|1\rangle) ~~~(where~~ j=0,1,2)\\
         |\beta_1^{0,j}\rangle=\frac{1}{\sqrt{2}}(|0\rangle-e^{-i\theta_j}|1\rangle),~~~(where~~ j=0,1,2).
     \end{split}
 \end{equation}
 When Alice's measurement outcomes are $|\alpha_1^j\rangle_{A_j}: j=0,1,2$ then Charlie select his measurement bases $\{|\beta_0^{1,j}\rangle, |\beta_1^{1,j}\rangle\}: j=0,1,2$ for measuring qubits $C_j:j=0,1,2$ respectively, which are given as
 \begin{equation}
     \begin{split}
         |\beta_0^{1,j}\rangle=&\frac{1}{\sqrt{2}}(e^{-i\theta_j}|0\rangle+|1\rangle) ~~~(where~~ j=0,1,2)\\
         |\beta_1^{1,j}\rangle=&\frac{1}{\sqrt{2}}(-e^{-i\theta_j}|0\rangle+|1\rangle).~~~(where~~ j=0,1,2)
     \end{split}
 \end{equation}
 At the same time, Bob, David and Candy make Bell-basis measurements on the qubit pairs $(1, B_1)$, $(2, D_0)$ and $(4, E_0)$, respectively. \\
 
\noindent Bell-basis is a two-qubit measurement basis consisting of four linearly independent sets of quantum states, which are given as 
 \begin{equation}
     \begin{split}
         |\Phi_{00}\rangle=\frac{1}{\sqrt{2}}(|00\rangle+|11\rangle),~~~~~& |\Phi_{01}\rangle=\frac{1}{\sqrt{2}}(|00\rangle-|11\rangle),\\
         |\Phi_{10}\rangle=\frac{1}{\sqrt{2}}(|01\rangle+|10\rangle),~~~~~& |\Phi_{11}\rangle=\frac{1}{\sqrt{2}}(|01\rangle-|10\rangle).
     \end{split}
 \end{equation}
 \textbf{Step 2:} Using all the above basis, we can write the state of the total system as
\begin{footnotesize}
 \begin{equation}
     \begin{split}
         |X\rangle=\frac{1}{\sqrt{2}}&\Bigg[\sum_{g_0=0}^{1}\frac{1}{\sqrt{2}}|\alpha_{g_0}^{0}\rangle_{A_0}\otimes\sum_{h_0=0}^{1}\frac{1}{\sqrt{2}}|\beta_{h_0}^{g_0,0}\rangle_{C_0}\otimes\big((-1)^{g_0}x_0|g_0\rangle+(-1)^{h_0}y_0e^{i\theta_0}|1\oplus g_0\rangle\big)_{B_0}\Bigg]\\
         \bigotimes&\Bigg[\sum_{g_1=0}^{1}\frac{1}{\sqrt{2}}|\alpha_{g_1}^{1}\rangle_{A_1}\otimes\sum_{h_1=0}^{1}\frac{1}{\sqrt{2}}|\beta_{h_1}^{g_1,1}\rangle_{C_1}\otimes\big((-1)^{g_1}x_1|g_1\rangle+(-1)^{h_1}y_1e^{i\theta_1}|1\oplus g_1\rangle\big)_{D_0}\Bigg]\\
         \bigotimes&\Bigg[\sum_{g_2=0}^{1}\frac{1}{\sqrt{2}}|\alpha_{g_2}^{2}\rangle_{A_2}\otimes\sum_{h_2=0}^{1}\frac{1}{\sqrt{2}}|\beta_{h_2}^{g_2,2}\rangle_{C_2}\otimes\big((-1)^{g_2}x_2|g_2\rangle+(-1)^{h_2}y_2e^{i\theta_2}|1\oplus g_2\rangle\big)_{E_0}\Bigg]\\
         \bigotimes&\Bigg[\sum_{m_0,n_0=0}^{1}\frac{1}{2}|\Phi_{m_0n_0}\rangle_{1B_1}\otimes\Big(a_0|m_0\rangle+(-1)^{n_0}b_0|1\oplus m_0\rangle\Big)_{A_3}\Bigg]\\
         \bigotimes&\Bigg[\sum_{m_1,n_1=0}^{1}\frac{1}{2}|\Phi_{m_1n_1}\rangle_{2D_1}\otimes\Big(a_1|m_1\rangle+(-1)^{n_1}b_1|1\oplus m_1\rangle\Big)_{A_4}\Bigg]\\
         \bigotimes&\Bigg[\sum_{m_2,n_2=0}^{1}\frac{1}{2}|\Phi_{m_2n_2}\rangle_{4E_1}\otimes\Big(a_2|m_2\rangle+(-1)^{n_2}b_2|1\oplus m_2\rangle\Big)_{A_5}\Bigg]\bigotimes|0\rangle_{S}
     \end{split}
 \end{equation}
 \end{footnotesize}
 \begin{footnotesize}
 \begin{equation*}
     \begin{split}
         +\frac{1}{\sqrt{2}}&\Bigg[\sum_{g_0=0}^{1}\frac{1}{\sqrt{2}}|\alpha_{g_0}^{0}\rangle_{A_0}\otimes\sum_{h_0=0}^{1}\frac{1}{\sqrt{2}}|\beta_{h_0}^{g_0,0}\rangle_{C_0}\otimes\big(x_0|g_0\rangle+(-1)^{g_0\oplus h_0\oplus 1}y_0e^{i\theta_0}|1\oplus g_0\rangle\big)_{B_0}\Bigg]\\
         \bigotimes&\Bigg[\sum_{g_1=0}^{1}\frac{1}{\sqrt{2}}|\alpha_{g_1}^{1}\rangle_{A_1}\otimes\sum_{h_1=0}^{1}\frac{1}{\sqrt{2}}|\beta_{h_1}^{g_1,1}\rangle_{C_1}\otimes\big(x_1|g_1\rangle+(-1)^{g_1\oplus h_1\oplus 1}y_1e^{i\theta_1}|1\oplus g_1\rangle\big)_{D_0}\Bigg]\\
         \bigotimes&\Bigg[\sum_{g_2=0}^{1}\frac{1}{\sqrt{2}}|\alpha_{g_2}^{2}\rangle_{A_2}\otimes\sum_{h_2=0}^{1}\frac{1}{\sqrt{2}}|\beta_{h_2}^{g_2,2}\rangle_{C_2}\otimes\big(x_2|g_2\rangle+(-1)^{g_2\oplus h_2\oplus 1}y_2e^{i\theta_2}|1\oplus g_2\rangle\big)_{E_0}\Bigg]\\
         \bigotimes&\Bigg[\sum_{m_0,n_0=0}^{1}\frac{1}{2}|\Phi_{m_0n_0}\rangle_{1B_1}\otimes\Big((-1)^{m_0}a_0|m_0\rangle+(-1)^{1\oplus m_0\oplus n_0}b_0|1\oplus m_0\rangle\Big)_{A_3}\Bigg]\\
         \bigotimes&\Bigg[\sum_{m_1,n_1=0}^{1}\frac{1}{2}|\Phi_{m_1n_1}\rangle_{2D_1}\otimes\Big((-1)^{m_1}a_1|m_1\rangle+(-1)^{1\oplus m_1\oplus n_1}b_1|1\oplus m_1\rangle\Big)_{A_4}\Bigg]\\
         \bigotimes&\Bigg[\sum_{m_2,n_2=0}^{1}\frac{1}{2}|\Phi_{m_2n_2}\rangle_{4E_1}\otimes\Big((-1)^{m_2}a_2|m_2\rangle+(-1)^{1\oplus m_2\oplus n_2}b_2|1\oplus m_2\rangle\Big)_{A_5}\Bigg]\bigotimes|1\rangle_{S},
     \end{split}
 \end{equation*}
 \end{footnotesize}
 After performing the measurements by every party(excluding controller Simon), they shared their measurement outcomes publicly via the classical channel. Suppose Alice's measurement results are $|\alpha_{g_0}^0\rangle_{A_0}$, $|\alpha_{g_1}^1\rangle_{A_1}$ and $|\alpha_{g_2}^2\rangle_{A_2}$, Charlie's measurement outcomes are $|\beta_{h_0}^{g_0,0}\rangle_{C_0}$, $|\beta_{h_1}^{g_1,1}\rangle_{C_1}$  and $|\beta_{h_2}^{g_2,2}\rangle_{C_2}$, Bob's outcome is $|\Phi_{m_0n_0}\rangle_{1B_1}$, David's outcome is $|\Phi_{m_1n_1}\rangle_{2D_1}$ and Candy's outcome is $|\Phi_{m_2n_2}\rangle_{4E_1}$ (here $g_j, h_j, m_j, n_j\in\{0,1\}$ for $j\in\{0,1,2\})$ then the reduced state of the system is given by
 
 \begin{equation}
     \begin{split}
         |R\rangle=&\frac{1}{\sqrt{2}}\big((-1)^{g_0}x_0|g_0\rangle+(-1)^{h_0}y_0e^{i\theta_0}|1\oplus g_0\rangle\big)_{B_0}
         \otimes\big((-1)^{g_1}x_1|g_1\rangle+(-1)^{h_1}y_1e^{i\theta_1}|1\oplus g_1\rangle\big)_{D_0}\\
         \otimes&\big((-1)^{g_2}x_2|g_2\rangle+(-1)^{h_2}y_2e^{i\theta_2}|1\oplus g_2\rangle\big)_{E_0}
         \otimes\big(a_0|m_0\rangle+(-1)^{n_0}b_0|1\oplus m_0\rangle\big)_{A_3}\\
         \otimes&\big(a_1|m_1\rangle+(-1)^{n_1}b_1|1\oplus m_1\rangle\big)_{A_4}
         \otimes\big(a_2|m_2\rangle+(-1)^{n_2}b_2|1\oplus m_2\rangle\big)_{A_5}\otimes|0\rangle_{S}\\
         +&\frac{1}{\sqrt{2}}\big(x_0|g_0\rangle+(-1)^{g_0\oplus h_0\oplus 1}y_0e^{i\theta_0}|1\oplus g_0\rangle\big)_{B_0}
         \otimes\big(x_1|g_1\rangle+(-1)^{g_1\oplus h_1\oplus 1}y_1e^{i\theta_1}|1\oplus g_1\rangle\big)_{D_0}\\
         \otimes&\big(x_2|g_2\rangle+(-1)^{g_2\oplus h_2\oplus 1}y_2e^{i\theta_2}|1\oplus g_2\rangle\big)_{E_0}
         \otimes\big((-1)^{m_0}a_0|m_0\rangle+(-1)^{1\oplus m_0\oplus n_0}b_0|1\oplus m_0\rangle\big)_{A_3}\\
         \otimes&\big((-1)^{m_1}a_1|m_1\rangle+(-1)^{1\oplus m_1\oplus n_1}b_1|1\oplus m_1\rangle\big)_{A_4}
         \otimes\big((-1)^{m_2}a_2|m_2\rangle+(-1)^{1\oplus m_2\oplus n_2}b_2|1\oplus m_2\rangle\big)_{A_5}\otimes|1\rangle_{S}.
     \end{split}
 \end{equation}
 \textbf{Step 3:} Controlled Simon examines all the measurement outcomes from all the others involved parties. If he satisfies, he measures his qubit $S$ with the basis $\{|0\rangle, |1\rangle\}$ and classically shares his measurement outcome $|t\rangle_{S} ($where $t\in\{0,1\})$ with all the parties.  \\

\noindent Finally, Alice, Bob, David, and Candy apply the unitary operators on their respective qubits to get the desired qubits to achieve the task.\\
 
\noindent Alice needs to apply the unitary operator on the qubits $(A_3, A_4, A_5)$ corresponding  to the measurement results 
  $|\Phi_{m_0n_0}\rangle_{1B_1}$, $|\Phi_{m_1n_1}\rangle_{2D_1}$, $|\Phi_{m_2n_2}\rangle_{4E_1}$ and $|t\rangle_{S}$ of Bob, David, Candy and Simon respectively, the unitary operators are given by\\
 \begin{equation}
     U_{Alice}= (1-t)U_{0}^{(m_j,n_j)}\otimes U_{1}^{(m_j,n_j)}\otimes U_2^{(m_j,n_j)}+t V_{0}^{(m_j,n_j)}\otimes V_{1}^{(m_j,n_j)}\otimes V_2^{(m_j,n_j)},
 \end{equation}
    where 
    \begin{align*}
        U_j^{(m_j,n_j)}=&|0\rangle\langle m_j|+(-1)^{n_j}|1\rangle\langle 1\oplus m_j|,\\
        V_j^{(m_j,n_j)}=&(-1)^{m_j}|0\rangle\langle m_j|+(-1)^{1\oplus m_j\oplus n_j}|1\rangle\langle 1\oplus m_j|.
    \end{align*}
    Now, Alice introduces three auxiliary qubits $A_6, A_7$ and $A_8$ within the initial state $0\rangle$ for each. and apply a CNOT gate on each of the qubit pairs $(A_4, A_6)$, $(A_5, A_7)$ and $(A_5, A_8)$ where $A_3$ and $A_5$ are the controlled qubit and qubits $A_6, A_7$ and $A_8$ are the targeted qubits. Then the state of the qubits $A_3, A_6, A_5, A_7$ and $A_8$ transformed into the state given by
   \begin{equation*}
     \begin{split}
(a_0|0\rangle+b_0|1\rangle)_{A_3}\otimes(a_1|00\rangle+b_1|11\rangle)_{A_4A_6}\otimes(a_2|000\rangle+b_2|111\rangle)_{A_5A_7A_8},
     \end{split}
 \end{equation*}
 Therefore, we see the three states that Bob, David and Candy want to teleport to Alice are successfully teleported into the qubits $A_3$, $(A_4A_6)$ and $(A_5A_7A_8)$, respectively.\\
 
 \noindent   Bob needs to apply the Unitary operator on his qubit $B_0$ to get the desired state corresponding to the measurement results of Alice, Charlie and Simon, $|\alpha^{0}_{g_0}\rangle_{A_0}$, $|\beta^{g_0,0}_{h_0}\rangle_{C_0}$ and $|t\rangle_{S}$ respectively, which is 
    \begin{equation}
     U_{Bob}= (1-t)M_{0}^{(g_0,h_0)}+t N_{0}^{(g_0,h_0)}.
 \end{equation}
 Then the state of the qubit $B_0$ becomes $(x_0|0\rangle+y_0e^{i\theta_0}|1\rangle)_{B_0}$ which Alice and Charlie jointly want to prepare in Bob's place.\\
 
\noindent David needs to apply the Unitary operator on his qubit $C_0$ to get the desired state corresponding to the measurement results of Alice, Charlie and Simon, $|\alpha^{1}_{g_1}\rangle_{A_1}$, $|\beta^{g_1,1}_{h_1}\rangle_{C_1}$ and $|t\rangle_{S}$ respectively, which is 
    \begin{equation}
     U_{David}= (1-t)M_{1}^{(g_1,h_1)}+t N_{1}^{(g_1,h_1)}.
 \end{equation}
 Then, the state of the qubit $D_0$ evolves $(x_1|0\rangle+y_1e^{i\theta_1}|1\rangle)_{D_0}$ which Alice and Charlie together desire to prepare in David's place.\\
 
 \noindent Candy needs to apply the Unitary operator on his qubit $C_0$ to get the desired state corresponding to the measurement results of Alice, Charlie and Simon, $|\alpha^{2}_{g_2}\rangle_{A_2}$, $|\beta^{g_2,2}_{h_2}\rangle_{C_2}$ and $|t\rangle_{S}$ respectively, which is 
    \begin{equation}
     U_{Candy}= (1-t)M_{2}^{(g_2,h_2)}+t N_{2}^{(g_2,h_2)},
 \end{equation} 
    where 
    \begin{align*}
        M_j^{(g_j,h_j)}=&(-1)^{g_j}|0\rangle\langle g_j|+(-1)^{h_j}|1\rangle\langle 1\oplus h_j|,\\
        N_j^{(g_j,h_j)}=&|0\rangle\langle g_j|+(-1)^{1\oplus g_j\oplus h_j}|1\rangle\langle 1\oplus m_j|.\\
    \end{align*}
    Then the state of the qubit $E_0$ becomes $(x_2|0\rangle+y_2e^{i\theta_2}|1\rangle)_{E_0}$ which Alice and Charlie jointly desire to prepare in Candy's place.\\
    
   \noindent  Therefore, regardless of the measurement outcomes for all parties, we see that Alice, Bob, David, and Candy always reconstructs their intended states. As a result, our protocol successfully fulfills its assignment, achieving a $100\%$ success probability or unit fidelity. 
  
  \section{Noise Analysis}\label{sec:noise}
 In this section, we would like to analyze the effect of noise on the average fidelity of the proposed protocol over two well-known Markovian channels, i.e., AD and PD. Suppose the controller 'Simon' generates the entangled states in his lab, distributes the qubits to respective parties through a noisy channel and keeps the requisite qubit $S$ with him.  Therefore, the qubits $S$ are not affected by the noisy environment, but all the remaining qubits traverse through noisy environments. Here, we consider the same noise for all the qubits at a time for simplification. \\

  \noindent  The density matrix of the whole initial state $|X\rangle$ of the entire system is given as
    \begin{equation}
        \rho=|X\rangle\langle X|.
    \end{equation}
 But specifically, in open quantum system formalism, a quantum state evolves under a noisy environment and the density matrix($\rho$) is transformed into the mixed state, which can be written in terms of Kraus operators as follows \cite{n1}
    \begin{equation}\label{eq17}
        \epsilon(\rho)=\sum_{j}E_{j}\rho E_{j}^{\dagger},
    \end{equation}
    where $E_j$ is the Kraus operator for the specific noisy channel.
    Now, all the involved parties make measurements on their respective qubits with corresponding measurement bases described in the main protocol. After the measurement is done, they share the classical information with one another. According to the classical information, they apply the unitary operation on the respective qubits. Then, the final reduced state of the qubits $B_0, D_0, E_0, A_3, A_4$ and $A_5$ is given by

    \begin{equation}\label{eq18}
        \rho_{out}=Tr_{123A_0C_0A_1C_1A_2C_2B_1D_1E_1S}(U\Theta \epsilon(\rho)\Theta^{\dagger}U^{\dagger}).
    \end{equation}
    where $\Theta=\Big(\bigotimes_{i=0}^{2}|\alpha_{g_i}^i\rangle_{A_i}\langle\alpha_{g_i}^i|\Big)\otimes\Big(\bigotimes_{i=0}^{2}|\beta_{h_i}^{g_i,i}\rangle_{C_i}\langle\beta_{h_i}^{g_i,i}|\Big)\otimes|\Phi_{m_0n_0}\rangle_{1B_1}\langle\Phi_{m_0n_0}|\otimes|\Phi_{m_1n_1}\rangle_{2D_1}\langle\Phi_{m_1n_1}|\otimes|\Phi_{m_2n_2}\rangle_{4E_1}\langle\Phi_{m_2n_2}|\otimes |t\rangle_{S}\langle t|\otimes I_{B_0D_0E_0A_3A_4A_5}$ and $U=U_{Alice}\otimes U_{Bob}\otimes U_{David}\otimes U_{Candy}\otimes I$. Here, the partial trace is taken over the qubits $1, 2, 3, A_0, C_0, A_1, C_1, A_2, C_2, B_1, D_1, E_1,$ and $ S$.\\
     Now, the fidelity of the process is defined as
     \begin{equation}\label{eq19}
         F=\langle\Psi_{in}  | \rho_{out}|\Psi_{in}\rangle.    
     \end{equation}
     Here the input state is given as $|\Psi_{in}\rangle =\bigotimes_{i=0}^{2}|\phi_{i}\rangle\bigotimes_{i=0}^{2}|\psi_{i}^{'}\rangle$.
    \subsection{Amplitude-Damping Noisy channel}\label{subsection:ad}
\begin{figure}[ht]
\begin{centering}
\includegraphics[scale=0.65]{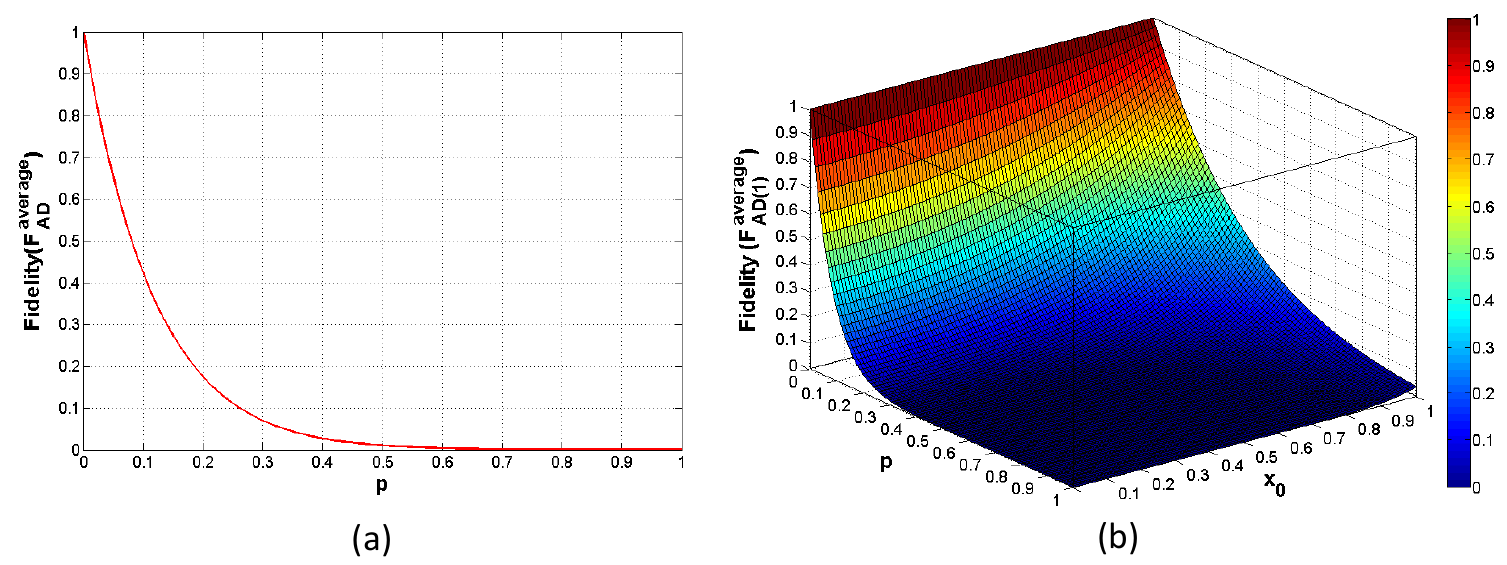}
\par\end{centering}
\centering
\caption{(Color online) (a) 2D plot shows the variation of $F_{AD}^{average}$ with respect to the noise parameter `$p$' when $x_0=\sqrt{0.3}, x_1=\sqrt{0.4}$ and $x_2=\sqrt{0.5}$. (b) 3D plot shows the variation of $F_{AD(1)}^{average}$ with respect to the noise parameter $p$ and the coefficient $x_0$.}
\label{fig:3}
\end{figure}

  Amplitude damping noise involves the energy dissipation in quantum systems and the Kraus operators for AD noisy channel are given as \cite{n1, adpd,n2,n3}
    \begin{equation}
        K_0=\begin{pmatrix}
            1&0\\0&\sqrt{1-p}
        \end{pmatrix},
        ~~~~~~~K_1=\begin{pmatrix}
            0&\sqrt{p}\\0&0
        \end{pmatrix}.
    \end{equation}
    where $p\in[0, 1]$ is the noise intensity or decoherence rate of the amplitude-damping channel.  Since we consider the effect of noise to be the same for all the qubits, the quantum channel given by the pure entangled $|\tau\rangle$ in (\ref{eq3}) is transformed according to linear mapping in (\ref{eq17}) into a mixed state which is given by the density matrix
    \begin{equation}
        \begin{split}
            \rho_{AD}=&(\bigotimes_{i=0}^{15} K_0)\otimes I|\tau\rangle\langle \tau| (\bigotimes_{i=0}^{15} K_0)^{\dagger}\otimes I+(\bigotimes_{i=0}^{15} K_1)\otimes I|\tau\rangle\langle \tau| (\bigotimes_{i=0}^{15} K_1)^{\dagger}\otimes I\\
            =&M\Big(|W_1^+\rangle_{A_0C_0B_0}\otimes|W_1^+\rangle_{A_1C_1D_0}\otimes|W_1^+\rangle_{A_2C_2E_0}\otimes|\Phi_{00}^{'}\rangle_{B_1A_3}\otimes|\Phi_{00}^{'}\rangle_{D_1A_4}\otimes|\Phi_{00}^{'}\rangle_{E_1A_5}\otimes|0\rangle_{S}\\
         +&|W_1^-\rangle_{A_0C_0B_0}\otimes|W_1^-\rangle_{A_1C_1D_0}\otimes|W_1^-\rangle_{A_2C_2E_0}\otimes|\Phi_{01}^{'}\rangle_{B_1A_3}\otimes|\Phi_{01}^{'}\rangle_{D_1A_4}\otimes|\Phi_{01}^{'}\rangle_{E_1A_5}\otimes|1\rangle_{S}\Big)\\
         \times&\Big(\langle W_1^+|_{A_0C_0B_0}\otimes\langle W_1^+|_{A_1C_1D_0}\otimes\langle W_1^+|_{A_2C_2E_0}\otimes\langle \Phi_{00}^{'}|_{B_1A_3}\otimes\langle \Phi_{00}^{'}|_{D_1A_4}\otimes\langle \Phi_{00}^{'}|_{E_1A_5}\otimes\langle 0|_{S}\\
         +&\langle W_1^-|_{A_0C_0B_0}\otimes\langle W_1^-|_{A_1C_1D_0}\otimes\langle W_1^-|_{A_2C_2E_0}\otimes\langle \Phi_{01}^{'}|_{B_1A_3}\otimes\langle \Phi_{01}^{'}|_{D_1A_4}\otimes\langle \Phi_{01}^{'}|_{E_1A_5}\otimes\langle 1|_{S}\Big)\\
         +&N|000000000000000\rangle_{A_0C_0B_0A_1C_1D_0A_2C_2E_0B_1A_3D_1A_4E_1A_5}\langle 000000000000000|\\ \otimes&(|0\rangle-|1\rangle)_S(\langle 0|-\langle 1|),
        \end{split}
    \end{equation}
    where 
    \begin{equation*}
    \begin{split}
        |W^{\pm}_1\rangle=\frac{1}{\sqrt{1+(1-p)^3}}&(|000\rangle\pm(\sqrt{1-p})^{3/2}|111\rangle),\\
        |\Phi_{00}^{'}\rangle=\frac{1}{\sqrt{1+(1-p)^2}}&(|00\rangle+(1-p)|11\rangle),\\
         |\Phi_{01}^{'}\rangle=\frac{1}{\sqrt{1+(1-p)^2}}&(|00\rangle-(1-p)|11\rangle),\\
         M=\frac{(1+(1-p)^3)^3(1+p^2)^3}{(1+(1-p)^3)^3(1+p^2)^3+p^{15}}&,~~~~ N=\frac{p^{15}}{(1+(1-p)^3)^3(1+p^2)^3+p^{15}}.
    \end{split} 
    \end{equation*}
      Alice, Charlie, Bob, David, Candy and Simon make their respective measurements on the respective qubits with the respective measurement bases. Then, they classically share their measurement results with the required parties and lastly, Alice, Bob, David and Candy apply a unitary operator on the respective qubits. Suppose Alice's measurement results are $|\alpha_{0}^0\rangle_{A_0}$, $|\alpha_{0}^1\rangle_{A_1}$ and $|\alpha_{0}^2\rangle_{A_2}$, Charlie's measurement outcomes are $|\beta_{0}^{0,0}\rangle_{C_0}$, $|\beta_{0}^{0,1}\rangle_{C_1}$  and $|\beta_{0}^{0,2}\rangle_{C_2}$, Bob's outcome is $|\Phi_{00}\rangle_{1B_1}$, David's outcome is $|\Phi_{00}\rangle_{2D_1}$,  Candy's outcome is $|\Phi_{00}\rangle_{4E_1}$ (here $g_j, h_j, m_j, n_j\in\{0,1\}$ for $j\in\{0,1,2\})$ and Simon's measurement result is $|0\rangle_{S}$ then after unitary is applied, the final reduced state according to (\ref{eq18}) is given by
      \begin{equation}
      \begin{split}
          \rho_{out}^{AD}=&M\big[\otimes_{i=0}^{2}(x_i|0\rangle+(1-p)^{3/2}y_i|1\rangle)\otimes(a_i|0\rangle+(1-p)b_i|1\rangle)\big]\times\big[\otimes_{i=0}^{2}(x_i\langle0|+(1-p)^{3/2}y_i\langle1|)\\
          &\otimes_{i=0}^{2}(a_i\langle0|+(1-p)b_i\langle1|)+N\bigotimes_{i=0}^{2}(x_i^2|0\rangle\langle 0|)\bigotimes_{i=0}^{2}(a_i^2|0\rangle\langle 0|).
          \end{split}
      \end{equation}
According to the equation (\ref{eq19}), the fidelity is given by
\begin{equation}
    F_{AD}=M\prod_{i=0}^{2}(x_i^{2}+(1-p)^{3/2}y_i^{2})^2\prod_{i=0}^{2}(a_i^{2}+(1-p)b_i^{2})^2+N\prod_{i=0}^{2}(x_i^{4})\prod_{i=0}^{2}(a_i^{4}),
\end{equation}
here $a_i=\cos(\frac{\theta_i}{2})$, $b_i=\sin(\frac{\theta_i}{2}) e^{i\xi_i}$. Now we calculate average fidelity which is obtained by taking the average over all unknown quantum states, i.e., by computing \cite {n2,n3}.
 \begin{equation}
 \begin{split}
    F_{AD}^{average}=M&\prod_{i=0}^{2}(x_i^{2}+(1-p)^{3/2}y_i^{2})^2\times\prod_{i=0}^{2}\Big(\frac{1}{4\pi}\int_{0}^{\pi}\sin{\theta}\int_{0}^{2\pi}(\cos^2{\frac{\theta_i}{2}}+(1-p)\sin{\frac{\theta_i}{2}})^2~d\xi d\theta\Big)\\
    +&N\prod_{i=0}^{2}(x_i^{4})\prod_{i=0}^{2}\Big(\frac{1}{4\pi}\int_{0}^{\pi}\sin{\theta}\int_{0}^{2\pi}\cos^4{\frac{\theta_i}{2}}~d\xi d\theta\Big)\\
    =M&\prod_{i=0}^{2}(x_i^{2}+(1-p)^{3/2}y_i^{2})^2\times \Big(\frac{3-3p+p^2}{3}\Big)^3+ \frac{N}{27}\prod_{i=0}^{2}(x_i^{4}).
    \end{split}
\end{equation}
The average fidelity depends on the noise intensity parameter `$p$' and the known states coefficients $(x_i,y_i)$.  When $x_0=\sqrt{0.3},  x_1=\sqrt{0.4}$ and $x_2=\sqrt{0.5}$  then the variation in average fidelity $F_{AD}^{average}$  with respect to the noise parameter `$p$' is given in figure \ref{fig:3} (a). If we consider that the three known states are the same, i.e., $x_0=x_1=x_2$,  then the average fidelity is given as
\begin{equation}
 \begin{split}
    F_{AD(1)}^{average}=\frac{M}{27}&\big(x_0^{2}+(1-p)^{3/2}(1-x_0^2)\big)^6\big(3-3p+p^2\big)^3+ \frac{N}{27}x_0^{12}.
    \end{split}
\end{equation}
and the variation of $F_{AD(1)}^{average}$ with respect to the noise parameter ($p$) and the coefficient $x_0$ is given in the figure \ref{fig:3} (b).

\noindent In AD noisy environment, it is evident from figure $2$ $(a)$ and $(b)$ that the average fidelity is decreasing with respect to an increase in the noise parameter `$p$'. Also, fidelity is near zero in the interval $(0.6, 1)$.

  \subsection{Phase-Damping Noisy channel}\label{subsection:pd}
Phase damping is the process of loss of information about the relative phases of a quantum state. In a phase-damping noisy channel, the Kraus operators are given as \cite{n1, adpd,n2,n3}
    \begin{equation}
        K_0=\begin{pmatrix}
            \sqrt{1-q}&0\\0&\sqrt{1-q}
        \end{pmatrix},
        K_1=\begin{pmatrix}
            \sqrt{q}&0\\0&0
        \end{pmatrix}, K_2=\begin{pmatrix}
            0&0\\0&\sqrt{q}
        \end{pmatrix},
    \end{equation}
    where $q\in[0, 1]$ represents the noise intensity of the phase-damping channel, which lies from $0$ to $1$.  Since we consider the effect of noise to be the same for all the qubits, the quantum channel given by the pure entangled $|\tau\rangle$ in (\ref{eq3}) transformed according to linear mapping in (\ref{eq17}) into a mixed state which is given by the density matrix
  \begin{equation}
      \begin{split}
           \rho_{PD}=&\sum_{j=0}^{2}\Big[(\prod_{i=0}^{15} K_j)\otimes I|\tau\rangle\langle \tau| (\prod_{i=0}^{15} K_j)^{\dagger}\otimes I\Big]\\
           =&\frac{(1-q)^{15}}{R}\Big(|W^+\rangle_{A_0C_0B_0}\otimes|W^+\rangle_{A_1C_1D_0}\otimes|W^+\rangle_{A_2C_2E_0}\otimes|\Phi_{00}\rangle_{B_1A_3}\otimes|\Phi_{00}\rangle_{D_1A_4}\otimes|\Phi_{00}\rangle_{E_1A_5}\otimes|0\rangle_{S}\\
         +&|W^-\rangle_{A_0C_0B_0}\otimes|W^-\rangle_{A_1C_1D_0}\otimes|W^-\rangle_{A_2C_2E_0}\otimes|\Phi_{01}\rangle_{B_1A_3}\otimes|\Phi_{01}\rangle_{D_1A_4}\otimes|\Phi_{01}\rangle_{E_1A_5}\otimes|1\rangle_{S}\Big)\\
         \times&\Big(\langle W^+|_{A_0C_0B_0}\otimes\langle W^+|_{A_1C_1D_0}\otimes\langle W^+|_{A_2C_2E_0}\otimes\langle \Phi_{00}|_{B_1A_3}\otimes\langle \Phi_{00}|_{D_1A_4}\otimes\langle \Phi_{00}|_{E_1A_5}\otimes\langle 0|_{S}\\
         +&\langle W^-|_{A_0C_0B_0}\otimes\langle W^-|_{A_1C_1D_0}\otimes\langle W^-|_{A_2C_2E_0}\otimes\langle \Phi_{01}|_{B_1A_3}\otimes\langle \Phi_{01}|_{D_1A_4}\otimes\langle \Phi_{01}|_{E_1A_5}\otimes\langle 1|_{S}\Big)\\
         +&\frac{q^{15}}{R}|000000000000000\rangle_{A_0C_0B_0A_1C_1D_0A_2C_2E_0B_1A_3D_1A_4E_1A_5}\langle 000000000000000|\\ \otimes&(|0\rangle+|1\rangle)_S(\langle 0|+\langle 1|)\\
         +&\frac{q^{15}}{R}|111111111111111\rangle_{A_0C_0B_0A_1C_1D_0A_2C_2E_0B_1A_3D_1A_4E_1A_5}\langle 111111111111111|\\ \otimes&(|0\rangle-|1\rangle)_S(\langle 0|-\langle 1|),
      \end{split}
  \end{equation}
  where $R=(1-q)^{15}+2q^{15}$ is the normalization factor.\\

\noindent Alice, Charlie, Bob, David, Candy, and Simon each conduct measurements on their respective qubits using specific measurement bases. Following this, they share their measurement outcomes with the necessary recipients. Subsequently, Alice, Bob, David, and Candy individually apply a unitary operator to their respective qubits. Let's denote Alice's measurement outcomes as $|\alpha_{0}^0\rangle_{A_0}$, $|\alpha_{0}^1\rangle_{A_1}$ and $|\alpha_{0}^2\rangle_{A_2}$, Charlie's  as $|\beta_{0}^{0,0}\rangle_{C_0}$, $|\beta_{0}^{0,1}\rangle_{C_1}$  and $|\beta_{0}^{0,2}\rangle_{C_2}$, Bob's as $|\Phi_{00}\rangle_{1B_1}$, David's as $|\Phi_{00}\rangle_{2D_1}$ and Candy's as $|\Phi_{00}\rangle_{4E_1}$ (here $g_j, h_j, m_j, n_j\in\{0,1\}$ for $j\in\{0,1,2\})$ and Simon's as $|0\rangle_{S}$. After the application of the unitary operation, the resulting final reduced state can be described 
 according to equation (\ref{eq18}) as:
\begin{equation}
    \begin{split}
          \rho_{out}^{PD}=&\frac{(1-q)^{15}}{R}\Big[\otimes_{i=0}^{2}(x_i|0\rangle+y_i|1\rangle)\otimes_{i=0}^{2}(a_i|0\rangle+b_i|1\rangle)\Big]\Big[\otimes_{i=0}^{2}(x_i\langle0|+y_i\langle1|)\otimes_{i=0}^{2}(a_i\langle0|+b_i\langle1|)\Big]\\
          &+\frac{q^{15}}{R}(\prod_{i-0}^2x_i^2a_i^2)|000000\rangle\langle 000000|+\frac{q^{15}}{R}(\prod_{i-0}^2y_i^2b_i^2)|111111\rangle\langle 111111|.
    \end{split}
\end{equation}
According to the equation (\ref{eq19}), the fidelity is given by
\begin{equation}
    F_{PD}=\frac{(1-q)^{15}}{R}+\frac{q^{15}}{R} \prod_{i=0}^{2}x_i^4a_i^4+\frac{q^{15}}{R}\prod_{i=0}^{2}y_i^4b_i^4.
\end{equation}
 We consider $a_i=\cos(\frac{\theta_i}{2})$, $b_i=\sin(\frac{\theta_i}{2}) e^{i\xi_i}$ where $\theta\in[0,\pi], \xi\in[0, 2\pi]$. Now, the average fidelity over all the unknown states is given by
 \begin{equation}
 \begin{split}
    F_{PD}^{average}=&\frac{(1-q)^{15}}{R}
    +\frac{q^{15}}{R}\prod_{i=0}^{2}(x_i^{4})\prod_{i=0}^{2}\Big(\frac{1}{4\pi}\int_{0}^{\pi}\sin{\theta}\int_{0}^{2\pi}\cos^4{\frac{\theta_i}{2}}~d\xi d\theta\Big)\\
    +&\frac{q^{15}}{R}\prod_{i=0}^{2}(y_i^{4})\prod_{i=0}^{2}\Big(\frac{1}{4\pi}\int_{0}^{\pi}\sin{\theta}\int_{0}^{2\pi}\sin^4{\frac{\theta_i}{2}}~d\xi d\theta\Big)\\
    =&\frac{(1-q)^{15}}{R}+\frac{q^{15}}{27R}\prod_{i=0}^{2}(x_i^{4})+\frac{q^{15}}{27R}\prod_{i=0}^{2}(y_i^{4}).
    \end{split}
\end{equation}

\begin{figure}[ht]
\begin{centering}
\includegraphics[scale=0.65]{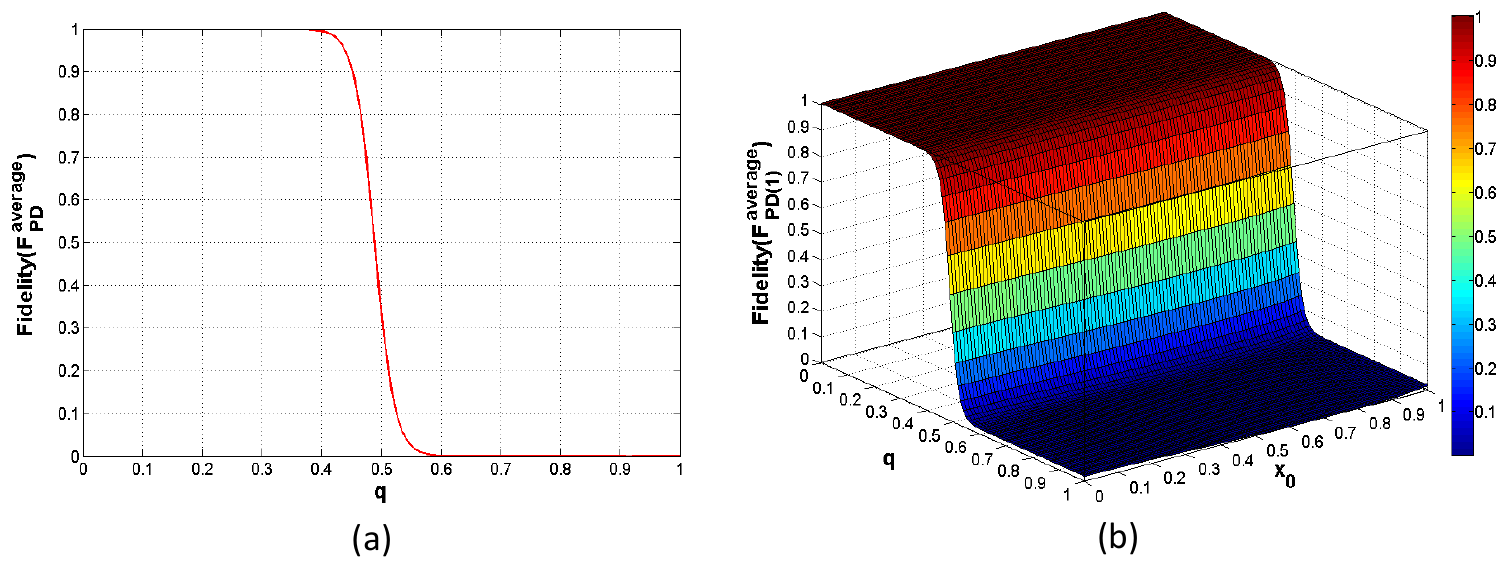}
\par\end{centering}
\centering{}
\caption{(Color online)(a) Variation of $F^{average}$ in the PD noisy channel with respect to the noise parameter `q' when the known state coefficients are $x_0=\sqrt{0.3}, x_1=\sqrt{0.4}$ and $x_2=\sqrt{0.5}$. (b) 3D plot shows the variation of $F_{PD(1)}^{average}$ with respect to noise parameter ($q$) and the coefficient $x_0$ .}
\label{fig:4}
\end{figure}

If three known states are the same, then the average fidelity is given as
\begin{equation}
 \begin{split}
    F_{PD(1)}^{average}=\frac{(1-q)^{15}}{R}+\frac{q^{15}}{27R}(x_0^{12}+y_0^{12}).
    \end{split}
\end{equation}
For PD noise, we consider the same known state coefficients as mentioned in the AD noisy channel (see \ref{subsection:ad}) and the variation of average fidelity under the PD noisy channel ($F_{PD(1)}^{average}$) concerning the noise parameter `q' and the coefficients are given in the figures \ref{fig:4} (a) and (b).

\noindent In PD noisy environment, it is evident that the average fidelity approaches unity within the range of $p\in[0, 0.4]$ in figures \ref{fig:4} (a) and (b) and diminishes towards zero within the range of $p\in[ 0.6, 1]$. Notably, the fidelity experiences a sharp decline within the interval $q\in (0.4, 0.6).$

\section{Discussion and Conclusion \label{sec:conclusion}} 
A hybrid six-party multi-directional CQT protocol is discussed in this paper with the use of a multi-qubit entangled quantum state. Especially, QT and JRSP quantum communication schemes are proposed in a protocol at the same time by using a 16-qubit entangled quantum state under the supervision of a controller 'Simon'. The HMCQC protocol is divided into a few steps to make it easier for new researchers to learn and implement, and it offers many benefits over the previous protocol. Several simple operations, including CNOT gate, single-qubit measurements, two-qubit measurements, Bell state measurements as well as unitary operations, have been used to achieve communication tasks successfully. Our work is not only limited to that, but we have also studied the protocol in an open quantum system and analytically studied the effect of AD and PD noises on the hybrid communication scheme by calculating average fidelity. We believe that hybrid multi-directional controlled quantum communication can effectively address the requirements of upcoming quantum communication networks and establish a solid basis for multi-party quantum communication.\\

\noindent \textbf{Acknowledgement} This work is supported by Indian Institute of Engineering Science and Technology, Shibpur, India. MS thankfully acknowledges the following funding agency: Defence Research and Development Organisation, India (JATC-2 PROJECT $\#$ 2(RP03700G)) for a project grant of  Prof. Joyee Ghosh (IIT Delhi, Department of physics) from where MS is drawing the salary while working with Prof. Binayak.\\

\end{document}